\documentclass[12pt]{article}
\usepackage{amssymb,epsf}
\def\lsim{\mathrel{\rlap {\raise.5ex\hbox{$ < $}}
{\lower.5ex\hbox{$\sim$}}}}
\def\gsim{\mathrel{\rlap {\raise.5ex\hbox{$ > $}}
{\lower.5ex\hbox{$\sim$}}}}
\def\sqr#1#2{{\vcenter{\vbox{\hrule height.#2pt
        \hbox{\vrule width.#2pt height#1pt \kern#1pt
           \vrule width.#2pt}
        \hrule height.#2pt}}}}

\def\lsim{{\displaystyle
{{\raise-8pt\hbox{$ <$}}
\atop{\raise5pt\hbox{$\sim$}}}}}
\def\gsim{{\displaystyle
{{\raise-8pt\hbox{$ >$}}
\atop{\raise5pt\hbox{$\sim$}}}}}
\def\slsim{{\displaystyle
{{\raise-8pt\hbox{$\scriptstyle <$}}
\atop{\raise5pt\hbox{$\scriptstyle \sim$}}}}}
\def\sgsim{{\displaystyle
{{\raise-8pt\hbox{$\scriptstyle  >$}}
\atop{\raise5pt\hbox{$\scriptstyle \sim$}}}}}
\newskip\humongous \humongous=0pt plus 1000pt minus 1000pt

\newcommand{\oao}[2]{{#1\atopwithdelims[]#2}}

\newcommand{\sump}[0]{\sum_{(h,g)}\!{\raise 4pt \hbox{$'$}}\,}
\newcommand{\sumpf}[0]{\sum_{(H^{\rm f},G^{\rm f})}\!\!\!\!{\raise 4pt
\hbox{$'$}}\,}
\newcommand{\sumpo}[0]{\sum_{(H^1,G^1)}\!\!\!\!{\raise 4pt
\hbox{$'$}}\,}
\newcommand{\sumpF}[0]{\sum_{(H^F,G^F)}\!\!\!\!{\raise 4pt
\hbox{$'$}}\,}

\def\bs{\begin{subequations}}
\def\es{\end{subequations}}
\catcode`\@=11
\newcount\hour
\newcount\minute
\newtoks\amorpm
\hour=\time\divide\hour by60
\minute=\time{\multiply\hour by60 \global\advance\minute by-\hour}
\edef\standardtime{{\ifnum\hour<12 \global\amorpm={am}%
        \else\global\amorpm={pm}\advance\hour by-12 \fi
        \ifnum\hour=0 \hour=12 \fi
        \number\hour:\ifnum\minute<10 0\fi\number\minute\the\amorpm}}
\edef\militarytime{\number\hour:\ifnum\minute<10 0\fi\number\minute}
\def\draftlabel#1{{\@bsphack\if@filesw {\let\thepage\relax
   \xdef\@gtempa{\write\@auxout{\string
      \newlabel{#1}{{\@currentlabel}{\thepage}}}}}\@gtempa
   \if@nobreak \ifvmode\nobreak\fi\fi\fi\@esphack}
        \gdef\@eqnlabel{#1}}
\def\@eqnlabel{}
\def\@vacuum{}
\def\draftmarginnote#1{\marginpar{\raggedright\scriptsize\tt#1}}
\def\draft{\oddsidemargin -.2truein
        \def\@oddfoot{\sl preliminary draft \hfil
        \rm\thepage\hfil\sl\today\quad\militarytime}
        \let\@evenfoot\@oddfoot \overfullrule 3pt
        \let\label=\draftlabel
        \let\marginnote=\draftmarginnote
   \def\@eqnnum{(\theequation)\rlap{\kern\marginparsep\tt\@eqnlabel}%
\global\let\@eqnlabel\@vacuum}  }
\def\subequations{\refstepcounter{equation}%
  \edef\@savedequation{\the\c@equation}%
  \@stequation=\expandafter{\theequation}%   %only want \theequation
  \edef\@savedtheequation{\the\@stequation}% % expanded once
  \edef\oldtheequation{\theequation}%
  \setcounter{equation}{0}%
  \def\theequation{\oldtheequation\alph{equation}}}

\def\endsubequations{\setcounter{equation}{\@savedequation}%
  \@stequation=\expandafter{\@savedtheequation}%
  \edef\theequation{\the\@stequation}\global\@ignoretrue
  \vspace*{-12pt} \\}
\def\bs{\begin{subequations}}
\def\es{\end{subequations}}
\relax
\def\Im{\,{\rm Im}\, }
\def\Re{\,{\rm Re}\, }

\def\bP{\overline{P}}

\def\bE{\overline{E}}

\def\bH{\overline{H}}
\def\bHt{\overline{H}_{\rm tw}}

\def\bOmega{\overline{\Omega}}

\def\g{\gamma}
\def\thefootnote{\fnsymbol{footnote}}
\def\be{\begin{equation}}
\def\ee{\end{equation}}
\def\ba{\begin{eqnarray}}
\def\ea{\end{eqnarray}}

\def\th{\vartheta}

\def\l{\lambda}

\def\d{\delta}
\def\g{\gamma}
\def\thet{\tau_{S_{\rm Het}}}
\def\t{\tau}

\def\t{\tau}
\def\im{\, {\rm Im}\, \tau}

\def\sp{\ , \ \ }
\def\ifd{\int_{\cal F}\frac{{\rm d}^2\tau}{\im}}
\newcommand{\ar}[2]{{#1\atopwithdelims[]#2}}

\def\ee{\end{equation}}
\def\bea{\begin{eqnarray}}
\def\eea{\end{eqnarray}}
\def\nn{\nonumber}
\def\nl{\hfil\break}

\def\np#1#2#3{Nucl. Phys. {\bf{B#1}} (#2) #3}
\def\pl#1#2#3{Phys. Lett. {\bf{B#1}} (#2) #3}

\newcommand{\Li}[0]{{\cal L}i}
\newcommand{\Lii}[1]{{{\cal L}i}_3\left(e^{2\pi
i\left(#1\right)}\right)}

\newcommand{\uarrw}[0]{\mathrel{
{\raise.5ex\vbox{\hrule width 1cm}\hskip-6pt\rightarrow}}}
\def\nl{\hfil\break}
\def\thebibliography#1{%
\vskip 0.5cm \centerline{\bf References}
\list{%
[\arabic{enumi}]}{\settowidth\labelwidth{[#1]}
\leftmargin\labelwidth
\advance\leftmargin\labelsep
\usecounter{enumi}}
\def\newblock{\hskip .11em plus .33em minus .07em}
\sloppy\clubpenalty4000\widowpenalty4000
\sfcode`\.=1000\relax}

\renewcommand{\theequation}{\arabic{section}.\arabic{equation}}

\renewcommand{\section}{\setcounter{equation}{0}\@startsection%
{section}{1}{0mm}{-\baselineskip}{0.5\baselineskip}%
{\normalfont\normalsize\bfseries}}

\renewcommand{\subsection}{\@startsection%
{subsection}{2}{0mm}{-\baselineskip}{0.5\baselineskip}%
{\normalfont\normalsize\slshape}}
\begin{document}
\renewcommand{\theequation}{\arabic{section}.\arabic{equation}}
\begin{titlepage}
\begin{flushright}
CERN-TH/98-407 \\ NEIP-98-019\\
LPTENS-98-46 \\ CPTH-S697.1298\\
hep-th/9901117\\
\end{flushright}
\begin{centering}
\vspace{.15in}
\boldmath
{\bf NON-PERTURBATIVE TRIALITY IN HETEROTIC}\\
{\bf AND TYPE II $N=2$ STRINGS$^{\ \dag}$}\\
\unboldmath
\vspace{.2in}
%\vspace{1. cm}
{Andrea GREGORI}$^{\ 1}$, {Costas KOUNNAS}$^{\ 2,\, \ast}$\\
\medskip
and \\
\medskip
{P. Marios PETROPOULOS}$^{\ 3}$\\
\vspace{.1in}
{\it $^1 $ Institut de Physique Th\'eorique, Universit\'e de
Neuch\^atel}\\
{\it 2000 Neuch\^atel, Switzerland.}\\
\medskip
{\it $^2 $ Theory Division, CERN}\\
{\it 1211 Geneva 23, Switzerland.}\\
\medskip
{\it $^3 $ Centre de Physique Th\'eorique,
Ecole Polytechnique}$^{\ \diamond}$\\
{\it 91128 Palaiseau Cedex, France.}\\
\vspace{.15in}

{\bf Abstract}\\
\end{centering}
The non-perturbative equivalence of four-dimensional $N=2$
superstrings with three vector multiplets and four hypermultiplets is
analysed. These models
are obtained through freely acting orbifold compactifications from the
heterotic, the symmetric and the asymmetric type II
strings. The heterotic
scalar manifolds are
$\big(SU(1,1) \big/ U(1)\big)^3$ for the $S,T,U$ moduli sitting
in the vector multiplets
and $SO(4,4)\big/ \big(SO(4) \times SO(4)\big)$ for those
in the hypermultiplets.
The  type II symmetric duals correspond to a self-mirror
Calabi--Yau threefold compactification with Hodge numbers
$h^{1,1}=h^{2,1}=3$, while the type II asymmetric construction
corresponds to
a spontaneous breaking of the $N=(4,4)$ supersymmetry to $N=(2,0)$.
Both have already been considered in the literature.
The heterotic construction instead is new and we
show that there is a weak/strong coupling  $S$-duality relation 
between the heterotic and
the asymmetric type IIA ground state with $4 \pi S_{\rm Het}=-(4 \pi
S_{\rm As})^{-1}$; we also show that there is a partial
restoration of $N=8$ supersymmetry
in the heterotic strong-coupling regime.
We compute the full (non-)perturbative $R^2$ and $F^2$ corrections
and determine the prepotential.
\vspace{.1in}
\begin{flushleft}
CERN-TH/98-407\\
January 1999\\
\end{flushleft}
\hrule width 6.7cm
\vskip.1mm{\small \small \small
$^\dag$\ Research partially supported by the EEC under the contracts
TMR-ERBFMRX-CT96-0045 and TMR-ERBFMRX-CT96-0090.\\
$^\ast$\ On leave from {\it Laboratoire de Physique Th\'eorique de
l'Ecole Normale Sup\'erieure,}
{\it CNRS,} 24 rue Lhomond, 75231 Paris Cedex 05, France.\\
$^\diamond$ Unit{\'e} mixte 7644 du {\it Centre National de la Recherche
Scientifique}.}
\end{titlepage}
\newpage
\setcounter{footnote}{0}
\renewcommand{\thefootnote}{\arabic{footnote}}

\section{Introduction}

In four space-time dimensions, string solutions
with $N \le 4$ supersymmetries can be constructed,
through appropriate compactifications of the
six-dimensional internal manifold, from
any of the  perturbative ten-dimensional strings:
heterotic, type I,  type IIA or type IIB. Although these
constructions
appear different at the string perturbative level, they might be
non-perturbatively equivalent, provided the massless
spectrum and the number of supersymmetries is the
same \cite{ht}--\cite{pw}.
As far as  $N=4$ supersymmetry is concerned, several tests in favour of
the non-perturbative
duality equivalence have been presented in the literature, not only in the
case of the heterotic string compactified on $T^6$ versus the type IIA, B
compactified on $K3\times T^2$\cite{ht},
but  also for theories with a lower number of massless vector
multiplets \cite{6auth}, including the type IIA, B $N=4$ asymmetric,
freely acting, orbifold constructions \cite{sv}.

In the $N=4$ theories  the space of the moduli fields is restricted
by supersymmetry to be the coset \cite{ht,fk}:
\be
\left({SL(2,R)\over U(1)}\right)_S
\times\left({SO(6,6+r)\over SO(6)\times
SO(6+r)}\right)_{T}\, ,
\ee
where $r=16$ in the heterotic and type IIA, B
compactified on  $T^6$ and $K3 \times T^2$, respectively.
Models with lower rank, $r<16$, can
be constructed, via freely acting (asymmetric) orbifold
compactification,
from any of the perturbative superstring theories in ten dimensions
\cite{6auth, fk,sv}.
On the heterotic side, the dilaton $S_{\rm Het}$ is always in the
gravitational multiplet, while on the type II side it is either
one of the moduli of the vector multiplets ($S_{\rm II}=T^1$),
when the  compactification is left--right-symmetric, with $N=(2,2)$
supersymmetry, or in the gravitational
multiplet in the asymmetric  compactification with $N=(4,0)$
supersymmetry.
The non-perturbative string duality therefore implies interchanges
between the moduli fields $S_{\rm Het}$ and $T^{i}$ of the scalar
manifold \cite{ht, 6auth, sv}, with the perturbative states of
one string theory mapped to non-perturbative states of its
dual equivalent and vice versa \cite{6auth, sv, s, hmn=4}.
The non-perturbative equivalence of dual strings has been verified
on several occasions: for instance, the anomaly cancellation of the
six-dimensional heterotic string implies that there should be a
one-loop correction to the gravitational $R^2$ term in the type II
theory.
Such a term was found by direct calculation in \cite{6auth, vw2}. Its
one-loop threshold correction, upon compactification to four
dimensions,
implies instanton corrections on the heterotic side, due to
five-branes being wrapped around the six-torus. 
Several other indications
are given for dual $N=4$ theories with rank $r<16$ \cite{6auth}.

Heterotic/type II dual pairs with lower supersymmetry, $N=2$,
share properties similar to those of $N=4$. In general
(non-freely-acting) symmetric orbifolds still give rise to $N=2$
heterotic/type II dual pairs in four dimensions \cite{kv}--\cite{re}. On
the heterotic side they can be interpreted as $K3$ plus gauge-bundle
compactifications, while on the type II side
they are Calabi--Yau compactifications of the ten-dimensional type
IIA theory. The heterotic dilaton is in a vector-tensor  multiplet, dual
to a vector, and the
vector moduli space receives both perturbative and non-perturbative
corrections. The hypermultiplet moduli space, however, does
not receive perturbative corrections; if $N=2$ is assumed to be unbroken,
it does not receive non-perturbative corrections either.
The dilaton in the  type II (symmetric) constructions is in a
tensor multiplet, dual to a
hypermultiplet, and the prepotential for the vector multiplets
receives only tree-level contributions. The tree-level type II
prepotential was computed and shown to give the correct one-loop
heterotic result. This provides a quantitative test of the duality
\cite{kv,re} and allows us to reach the non-perturbative corrections
of the heterotic side.

In the quantitative tests of non-perturbative
dualities, extended supersymmetry plays an essential role, since it
allows for the existence of BPS states (states in short representations
of the supersymmetry algebra). These states are (generically)
non-perturbatively stable and provide a reliable window into
non-perturbative corrections to some terms of the effective action
that receive  contributions only from those states. The relevant
structures for this analysis are the helicity supertrace formulae,
which
distinguish between various BPS and non-BPS states
\cite{6auth, bk}--\cite{n=6}. For $N \ge 2$,
these  supertraces appear in particular
in the $F^2$ and $R^2$ (two-derivative) terms or in a special class
of higher-order terms constructed out of the Riemann tensor and the
graviphoton field strength \cite{fg}. In the four-dimensional
heterotic string, these terms are anomaly-related and it can be shown
that they receive only tree- and one-loop corrections. In higher
dimensions, they do not receive non-perturbative corrections
either \cite{bk2}.

The non-perturbative equivalence of some
heterotic/type II dual pairs with $N=2$ supersymmetry has been
investigated
in Refs. \cite{sv,fhsv,hmn=2,gkp}. In this class of $N=2$ models the
scalar manifolds are  coset spaces:
\be
{SU(1,1)\over U(1)}\times {SO(2,2+N_V)\over SO(2)\times SO(2+N_V)}
\ \ {\rm and}
\ \ {SO(4,4+N_H)\over SO(4)\times SO(4+N_H)},
\ee
describing the moduli space of the $N_V+3$ moduli in vector
multiplets and the $N_H+4$ in the hypermultiplets, respectively.
The type II symmetric duals correspond to
self-mirror Calabi--Yau threefold compactifications with Hodge
numbers $h^{1,1}= N_V +3=h^{2,1}=N_H + 3$, which are $K3$ fibrations,
necessary condition for the existence of heterotic duals
\cite{klm,al}.  The equivalence of heterotic/type II
model(s) with $N_V=N_H=8$ was studied in
Refs. \cite{fhsv,hmn=2,gkp}; recently, in Ref. \cite{gkp}, this
analysis has been
extended to type II and heterotic duals with $N_{V}=N_{H}=4$ and
2.

The purpose of this work is to extend the analysis of Ref. \cite{gkp}
and establish  the non-perturbative equivalence  of three different
$N=2$ constructions with $N_V=N_H=0$: the heterotic construction with
supersymmetry
$N=(2,0)$,  the symmetric type II construction with $N=(1,1)$ and
the asymmetric type II with $N=(2,0)$ supersymmetry.
All these constructions are based on six-dimensional
freely acting (asymmetric) orbifold compactifications and thus the
initial $N=8$ (in type II) or $N=4$ (in heterotic) supersymmetry is
spontaneously broken to $N=2$ \cite{kk}.
The heterotic scalar manifold of this $N=2$ construction is
described by the vector scalar manifold $\big(SU(1,1) \big/
U(1)\big)^3$
associated to  the three moduli $S,T$ and $U$, and by the
hypermultiplet
quaternionic one, $SO(4,4)\big/ \big(SO(4) \times SO(4)\big)$.
The  type IIA symmetric construction corresponds to a self-mirror
Calabi--Yau threefold compactification with Hodge numbers
$h^{1,1}=h^{2,1}=3$ \cite{sv}. The type II asymmetric construction
\cite{sv}
corresponds to
a spontaneous breaking of $N=(4,4)$ to $N=(2,0)$
supersymmetry \cite{kk}.

In Section 2 we construct the symmetric type II model with $N=(1,1)$
supersymmetry and calculate the one-loop gravitational  corrections
associated to the  $R^2$ term. The asymmetric type II
$N=(2,0)$ construction
is presented in Section 3; in the same section we show that the
$R^2$ corrections in the two type II dual theories
are in agreement with their non-perturbative equivalence.
The heterotic construction, as well as the corresponding corrections to
the gauge and gravitational couplings, are presented in Section
4. In Section 5 we compare the  heterotic, the type II symmetric and the
type II asymmetric corrections, and show that the non-perturbative
equivalence of the three models is verified; we
furthermore show that due to this triality equivalence, there
exists a  weak--strong coupling relation ($S$-duality) between the
heterotic and the asymmetric type II theory 
($4 \pi S_{\rm Het}=-(4 \pi S_{\rm As})^{-1}$). We also claim 
that the $N=8$ supersymmetry is restored in the heterotic strong
coupling regime. Finally, in Section 6 and in the appendix,
we derive  the perturbative 
prepotential, as well as part of its non-perturbative corrections, which 
are argued to be valid for all three dual theories.
Our conclusion and comments are given in Section 7.

\boldmath
\section{The $N=(1,1)$ type II symmetric
construction}
\unboldmath

We start by considering a type II symmetric construction with
$N=(1,1)$ supersymmetry and  no vector multiplets or hypermultiplets in
the twisted sector.
This model is obtained by compactification of the 
ten-dimensional type II string on a Calabi--Yau manifold
${\rm CY}^{(3,3)}$,
with Hodge numbers $h^{1,1}=h^{2,1}=3$. This compactification
reduces the $N=(4,4)$ supersymmetry
to the desired $N=(1,1)$.
In what follows  we will always work at the $Z_2^{(1)}\times
Z_2^{(2)}$
(freely acting) orbifold limit of this manifold, where the partition
function and the one-loop gravitational and gauge corrections
can be computed analytically.

At the orbifold point, where
${\rm CY}^{(3,3)}\equiv  T^6\bigg / \left(Z_2^{(1)}\times
Z_2^{(2)}\right)$,
the partition function of the model can be
written easily in terms of the characters of the  twisted and shifted
compactified left  and right coordinates $X^I, \bar X^I,I=1,\ldots,6$,
and in terms of twisted fermionic superpartners $\Psi^I$ and 
${\bar\Psi}^I$.
The remaining contribution to the partition function comes from the
left- and right-moving  non-compact supercoordinates
$X^{\mu}, \Psi^{\mu}, {\bar X}^{\mu}, {\bar \Psi}^{\mu}$
and the super-reparametrization ghosts
$b,c,\beta,\gamma$ and ${\bar b},{\bar c},{\bar \beta},{\bar
\gamma}$.
The resulting partition function reads:
\ba
Z_{\rm II}^{(1,1)} & = &
{1 \over \Im \tau \vert \eta \vert^{24} } {1 \over 4}
\sum_{H^1,G^1} \sum_{H^2,G^2}
\Gamma_{6,6} \ar{H^1,H^2}{G^1,G^2} \nonumber \\
&&\nonumber \\
&& \times {1 \over 2} \sum_{a,b} (-)^{a+b+ab} \vartheta \ar{a}{b}
\vartheta \ar{a+H^2}{b+G^2}
\vartheta \ar{a+H^1}{b+G^1}
\vartheta \ar{a-H^1-H^2}{b-G^1-G^2} \nonumber \\
&&\nonumber \\
&& \times {1 \over 2} \sum_{\bar{a},\bar{b}}
(-)^{\bar{a}+\bar{b}+\bar{a}\bar{b}} \bar{\vartheta}
\ar{\bar{a}}{\bar{b}}
\bar{\vartheta} \ar{\bar{a}+H^2}{\bar{b}+G^2}
\bar{\vartheta} \ar{\bar{a}+H^1}{\bar{b}+G^1}
\bar{\vartheta} \ar{\bar{a}-H^1-H^2}{\bar{b}-G^1-G^2}\, ,
\label{zII}
\ea
where the contribution of $\beta,\gamma, \Psi^{\mu},\Psi^{I}$ and
${\bar \beta}, {\bar \gamma}, {\bar \Psi}^{\mu},{\bar \Psi}^{I}$
gives rise to the functions $\vartheta$ and $\bar{\vartheta}$, while
$\Gamma_{6,6} \ar{H^1,H^2}{G^1,G^2}$ denotes
the contribution of  $X^I$ and $\bar X^I$; $(H^{1},G^{1})$ refer to
the boundary conditions
introduced by the
projection $Z_2^{(1)}$ and $(H^{2},G^{2})$ to the
projection $Z_2^{(2)}$:
\be
\Gamma_{6,6} \ar{H^1,H^2}{G^1,G^2}=
\Gamma_{2,2}^{(1)} \ar{H^2 \vert H^1}{G^2 \vert G^1}\,
\Gamma_{2,2}^{(2)} \ar{H^1 \vert H^1+H^2}{G^1 \vert G^1+G^2}\, 
\Gamma_{2,2}^{(3)} \ar{H^1+H^2 \vert H^2}
{G^1+G^2 \vert G^2}\, .
\ee
Here we have introduced the twisted and shifted
characters of a $c=(2,2)$ block,
$\Gamma_{2,2} \ar{h\vert h'}{g\vert g'}$; the first column refers to
the twist, the second to the shift. The non-vanishing components
are the following:
\ba
\Gamma_{2,2} \ar{h\vert h'}{g\vert g'}
&=&
{4\,  \vert\eta \vert^6\over
\left\vert
\vartheta{1+h\atopwithdelims[] 1+g}\,
\vartheta{1-h\atopwithdelims[] 1-g}
\right\vert} \sp
{\rm for\ } (h',g')=(0,0) \
{\rm or\ } (h',g')=(h,g) \nn \\
&=&
\Gamma_{2,2} \ar{h'}{g'} \sp
{\rm for\ } (h,g)=(0,0)\, ,
\label{g22ts}
\ea
where $\Gamma_{2,2} \ar{h'}{g'}$ is the $Z_2$-shifted $(2,2)$ lattice
sum. As usual, the shift has to be specified by
the way it acts on the   momenta  and windings (our conventions are
those of Refs. \cite{6auth,gkp,kkprn}).
Since the three complex planes of $T^6$ are translated,
there are no fixed points. Therefore there are no
extra massless states coming from the twisted sectors:
the massless spectrum of this model contains
the $N=2$ supergravity multiplet, 3 vector multiplets,
1 tensor multiplet and 3 hypermultiplets.
The tensor multiplet is the type II dilaton multiplet
and is equivalent to an extra hypermultiplet.

By using the techniques developed in Refs. \cite{kk, solving, kkprn},
it can be shown that this model possesses an $N=8$ supersymmetry
spontaneously broken through a super-Higgs phenomenon,
due to the free actions of $Z_2^{(1)}\times Z_2^{(2)}$.
There exist
appropriate limits of the moduli, which depend on the precise shifts
in the lattices, in which there
is an approximate restoration of 16 or 32 supercharges.
In such limits, the supersymmetry restoration
is accompanied by a logarithmic instead of a linear blow-up of
the various thresholds.
The logarithmic blow-up is an infrared artefact, which can be lifted
by switching on an infrared cut-off
$\mu$ larger than the mass of the extra massive gravitinos; the
thresholds  thus vanish, as expected, in the limit  ${m_{3/2}/ \mu} \to
0$ in which supersymmetry is extended to $N=8$.

The relevant quantities for the computation of the
string correction to the $R^2$ term are the helicity
supertraces $B_{2n}$.
These are defined as the vacuum expectation value
of $\left( Q+\overline{Q} \right)^{2n}$,
where $Q$, $\overline{Q}$ stand for the left- and
right-helicity contributions to the four-dimensional physical helicity.
For the details of the computation
of such quantities in the framework of the above models
we refer to previous publications \cite{gkp, kkprn}. Here we simply
quote the results.
A straightforward computation shows that $B_2 = 0$, as expected in
the models with $N_V=N_H$ \cite{gkp}.
This feature is common to all the
$N=2$ type II $Z_2 \times Z_2$ symmetric orbifolds \cite{gkr}, in which
$B_2$ can receive a non-zero contribution only from the
$N=(1,1)$ sectors of the orbifold. The
internal coordinates in these sectors are twisted; all corrections
are therefore moduli-independent and come from the massless
states only. One finds $B_2=B_2\vert_{\rm massless}$, which vanishes
in the model we are considering here.

On the
other hand, $B_4$  receives non-zero contributions from the $N=(2,2)$
sectors of the orbifold, and we find\footnote{The prime summation
over $(h,g)$ stands for $(h,g) =
\{(0,1),(1,0),(1,1)\} $.}:
\be
B_4=6  \sum_{i=1,2,3} \sump
\Gamma_{2,2}^{(i)}\ar{0|h}{0|g}\, .
\label{B411}
\ee
From (\ref{B411}),
by applying the techniques developed in \cite{kkprn},
one can see that there is a limit
in moduli space in which  $B_4$ vanishes; this is the signal of
the restoration of the $N=8$ supersymmetry.

The four-derivative gravitational correction we consider here
is similar to those that were analysed in Refs. \cite{6auth} and
\cite{gkp};
in order to obtain it we proceed as in \cite{gkp}.
There is no tree-level contribution to this operator, and the $R^2$
correction appears at one loop; it is related to the contribution of
the
$h^{1,1}$ moduli, obtained through the insertion, in the
one-loop partition function,
of the two-dimensional operator $2Q^2 \overline{Q}^2$.
In this class of models the
contribution of the $N=(1,1)$ sectors to $B_4$ vanishes, and
therefore
$\left\langle2Q^2 \overline{Q}^2 \right\rangle$ is $B_4 / 3$.
The massless contributions of the latter
give rise to an infrared logarithmic behaviour 
${B_4 \vert_{\rm massless}\over 3} 
\log \left( M^{(\rm IIA)\,2} \Big/
                       \mu^{(\rm IIA)\,2}\right)$
\cite{infra,delgrav}, where
$M^{(\rm IIA)}\equiv 1/\sqrt{\alpha'_{\rm IIA}}$ is
the type IIA string scale and $\mu^{(\rm IIA)}$ is the type IIA
infrared cut-off.
Besides this running, the one-loop correction contains, as usual, the
thresholds $\Delta_{\rm IIA}$, which account for the infinite tower of
massive string modes.
The threshold corrections to the $R^2$ term are related to the
infrared-regularized genus-one integral of $B_4 / 3$. This 
relationship can be made more precise by noting that the amplitude
$\left\langle2Q^2 \overline{Q}^2 \right\rangle$ contains more than
the $R^2$-term corrections: it also accounts for terms
such as $F^2$ or $H^2$. However, in the type IIA
string, the $R^2$ corrections depend on the K\"ahler moduli 
$T^1, T^2$ and $T^3$
(spanning the vector manifold), and are independent of the
complex-structure moduli $U^1, U^2$ 
and $U^3$ (spanning the scalar manifold) \cite{6auth, gkp}. We 
thus have
\be
\partial_{T^i}\Delta_{\rm IIA}=\frac{1}{3}\ifd
\partial_{T^i} B_4
\ , \ \
\partial_{U^i}\Delta_{\rm IIA}=0\, .
\label{IIAthr}
\ee
For definiteness
we choose the half-unit shifts for $Z_2^{(1)}$ and $Z_2^{(2)}$ 
as defined by the following insertions:
$(-)^{n_2G^1}$, $(-)^{m_1(G^1+G^2)}$, $(-)^{n_2G^2}$ shifting
the lattices of the first, second and third plane, respectively.
With this choice of lattice shifts the one-loop-corrected
gravitational coupling reads (up to a constant):
\be
{16 \, \pi^2\over g^2_{\rm grav}(\mu^{(\rm IIA)})}  =
- 2 \sum_{i=1,3} \log \Im T^i
\left\vert \th_2\left(T^i\right)
\right \vert^4
- 2\log \Im T^2
\left\vert \th_4\left(T^2\right)
\right \vert^4 + 6 \log {M^{(\rm IIA)}\over \mu^{(\rm IIA)}}
\, . \label{thrint}
\ee
The shifts on the $\Gamma^{(i)}_{2,2}$ lattices break
the $SL(2,Z)_{T^i}$ duality groups, and the actual subgroup
left unbroken depends on the kind of shifts performed
(see Refs. \cite{6auth, gkp, solving, kkprn}).
Furthermore, there are three $N=4$ restoration limits,
corresponding to  $\left(\Im T^1, 1 / \Im T^2\right) \to 0$,
$\left(\Im T^1, \Im T^3\right) \to 0$ or
$\left(1/ \Im T^2, \Im T^3\right) \to 0$.
The masses of the three extra pairs of gravitinos are in fact given by
\ba
m^2_{3/2}(1) & = & { 1 \over 4} \Im T^1 \Im U^1+
{ 1 \over 4} {\Im U^2 \over \Im T^2} \nn\\
m^2_{3/2}(2) & = & { 1 \over 4} \Im T^1 \Im U^1+
{ 1 \over 4} \Im T^3 \Im U^3 \\
m^2_{3/2}(3) & = & { 1 \over 4} {\Im U^2 \over \Im T^2}+
{ 1 \over 4} \Im T^3 \Im U^3 \, ,\nn
\ea
and each of them vanishes in one of the above limits.
Owing to the effective restoration of the $N=4$ supersymmetry,
there is no linear divergence in the volume of the decompactifying
manifold.
For example, when $\left(1/ \Im T^2, \Im T^3\right) \to 0$, we
observe the following leading behaviour:
\be
{16 \, \pi^2\over g^2_{\rm grav}(\mu^{(\rm IIA)})}
\to
-2\log \Im T^2 + 2 \log \Im T^3 \, .
\label{logT}
\ee
\nl
However, the threshold correction blows up linearly
with respect to the modulus of the first plane in the limit  $\Im T^1
\to \infty$:
\be
{16 \, \pi^2\over g^2_{\rm grav}(\mu^{(\rm IIA)})} \to 8\pi \Im T^1
\, .
\label{linear}
\ee
Finally, the $N=8$ supersymmetry is restored  when
$\left(\Im T^1, 1/ \Im T^2, \Im T^3\right) \to 0$. In this limit, the
correction
behaves logarithmically in all three moduli:
\be
{16 \, \pi^2\over g^2_{\rm grav}(\mu^{(\rm IIA)})}\to
 2 \log \Im T^1- 2 \log \Im T^2 +2 \log \Im T^3
\, .
\ee

\boldmath
\section{The $N=(2,0)$ type II asymmetric construction}
\unboldmath

We now consider the asymmetric type II orbifold, which is obtained
from the $N=8$ IIA superstring compactified on $T^6$
by applying the freely acting projections $Z_2^{ F_{\rm R}}$ and
$Z_2^{(1)}$. The latter is the same projection we considered
in the previous section:
it acts as a combination of rotation and translation, and it
reduces symmetrically the number of  supersymmetries by one half. Instead,
$Z_2^{ F_{\rm R}}$ acts as $(-)^{F_{\rm R}}$ together
with a translation on $T^6$, and
projects out all the right-moving supersymmetries.
The properties of the $N=4$ model obtained by applying only $Z_2^{
F_{\rm R}}$
were already analysed in \cite{6auth}.
The orbifold obtained by further application of $Z_2^{(1)}$
has an $N=(2,0)$ supersymmetry  realized  among the left-movers only.

The partition function of the model reads:
\ba
Z_{\rm II}^{2,0} & =  & {1 \over \Im \tau |\eta|^{24} }
{1 \over 4} \sum_{H^1,G^1} \sum_{H^{F},G^{F}}
\Gamma_{6,6} \ar{H^1, H^F}{G^1, G^F} \nonumber \\
&& \times {1 \over 2}
\sum_{a,b}(-)^{a+b+ab}
\vartheta^2 \ar{a}{b}
\vartheta \ar{a+H^1}{b+G^1}\vartheta \ar{a-H^1}{b-G^1} \nonumber \\
&& \times {1 \over 2}
\sum_{\bar{a},\bar{b}}(-)^{\bar{a}+\bar{b}+\bar{a}\bar{b}}
(-)^{\bar{a}G^{ F}+\bar{b}H^{ F}+H^{ F}G^{F}}
\bar{\vartheta}^2 \ar{\bar{a}}{\bar{b}}
\bar{\vartheta} \ar{\bar{a}+H^1}{\bar{b}+G^1}
\bar{\vartheta} \ar{\bar{a}-H^1}{\bar{b}-G^1}\, ,
\label{z20}
\ea
where now
\be
\Gamma_{6,6} \ar{H^1, H^F}{G^1, G^F} =
\Gamma_{2,2}^{(1)} \ar{0|H^1}{0|G^1}
\Gamma_{2,2}^{(2)} \ar{H^1|H^{ F}}{G^1|G^{ F}}
\Gamma_{2,2}^{(3)} \ar{H^1|H^1}{G^1|G^1}\, .
\label{g6620}
\ee

The massless spectrum contains, besides the supergravity multiplet,
1 vector-tensor multiplet, dual to a vector,
2 vector multiplets and 4 hypermultiplets: it is therefore the same
as that of the type II symmetric orbifold. However, there is
an important difference in the nature of the fields:
in this case the dilaton belongs to a vector multiplet.
This is a general property of all $N=(2,0)$ string
compactifications,
where supersymmetry charges involve left-movers only.
The reason is that
the dilaton, in such cases, is uncharged under the
$SU(2)$ operators that rotate the supercharges of the $N=2$
supergravity.

The three vector moduli are in this case the dilaton $S_{\rm As}$,
the K\"{a}hler class $T_{\rm As}$, and the complex structure
$U_{\rm As}$ of the first
torus. When $(H^1,G^1)=(0,0)$, expressions (\ref{z20}) and
(\ref{g6620}) give half the partition function of an $N=(4,0)$
asymmetric orbifold with gauge group $U(1)^6$.
This model was analysed in detail in Ref. \cite{6auth}.

By using the same techniques as for the type II symmetric orbifold,
it can be shown that the model at hand possesses a spontaneously broken
$N=8$ supersymmetry, due to the free action of $Z_2^{(1)}$
and $Z_2^{ F_{\rm R}}$.
This can be seen again from the analysis of the helicity supertraces.
We find that $B_2$ is a constant also
in this asymmetric construction. There are therefore
no ``$N=2$ singularities'', i.e. lines in moduli space 
with enhancement of the massless spectrum such that
$\Delta N_V \neq \Delta N_H$.
On the other hand, for finite values of the moduli,
there are no ``$N=4$ singularities'' either
(lines where $\Delta N_V= \Delta N_H$), 
because the bosonic vacuum energy
is $-1/2$, and there are no points in moduli space in which
new massless states can appear.

The helicity supertrace 
$B_4$ receives non-zero contributions from three $N=4$ sectors:
the $N=(4,0)$
sector with $(H^1,G^1)=(0,0)$,
the $N=(2,2)$ sector with
$(H^1,G^1)\ne (0,0)$ and $(H^{ F}, G^{ F})=(0,0)$, and 
the $N=(2,2)$ sector with $(H^1,G^1)\ne (0,0)$ and $(H^{ F},
G^{ F})=(H^1,G^1)$. We obtain:
\ba
B_4 & = & { 3 \over 8} {1 \over {\bar\eta}^{12}}
\sumpF (-)^{H^{ F} G^{ F}}
{\bar \vartheta}^4 \ar{1-H^F}{1-G^F} \Gamma_{6,6} \ar{0,H^{
F}}{0,G^{ F}}
\nonumber \\
&& + 12 \sumpo
\Gamma_{2,2}^{(1)} \ar{0|H^1}{0|G^1}.
\label{B420}
\ea
The contributions of the first line come from the $N=(4,0)$ sector,
while those of the second line are due to the $N=(2,2)$
sectors.

Expression (\ref{B420}) has to be compared with the analogous for
the type II symmetric orbifold, Eq. (\ref{B411}).
In both cases $B_4 \vert_{\rm massless}=18$, in agreement with
the expected $N=2$ supergravity result.
As in the type II symmetric orbifold, we can make  $B_4$
vanish by taking  appropriate limits in the
space of moduli belonging to vector multiplets and hypermultiplets.
In the asymmetric type II the vector-multiplet moduli
are the moduli  of the first complex plane $T_{\rm As},U_{\rm As} $.
The moduli of second and third complex planes belong to the
hypermultiplet space.
The lattice sum in the second line of (\ref{B420}) vanishes
in some  appropriate limits in $T_{\rm As}$ and $U_{\rm As}$. However,
only by taking further limits in some of the moduli
belonging to hypermultiplets can we make
also $\Gamma_{6,6}\ar{0,H^F}{0,G^F}$ vanish. This is precisely
the limit
in which the extra massive gravitinos of the asymmetric construction
become massless.

As was already pointed out in the framework of symmetric type II
constructions,  the $R^2$ gravitational corrections of the asymmetric
case do not receive any contribution beyond one loop. These
corrections are related  to the insertion of the operator 
$2Q^2 \overline{Q}^2$. Again, this amplitude accounts for more
terms (like $H^2$) and only half of it is relevant to the $R^2$.
Therefore, the only non-zero contribution 
is provided by one sixth of the $N=(2,2)$ sectors of $B_4$ (the second
term in Eq. (\ref{B420})). The part of $B_4$ associated
with the  $N=(4,0)$ sector does not enter in the $R^2$ correction and
thus the
moduli dependence comes from the vector multiplets only; there is no
dependence at all on  the hypermultiplet moduli, as
expected from general properties of the $N=2$ theories.
Both moduli $T_{\rm As}$ and $U_{\rm As}$
of the first plane belong to vector multiplets and appear in
the correction to the $R^2$ term.

With the specific choice of half-unit shift, $(-)^{m_1G^1}$,
induced by $Z_2^{(1)}$ acting on $\Gamma_{2,2}^{(1)}$, we obtain
the corrected gravitational coupling in terms of the moduli 
$T_{\rm As}$, $U_{\rm As}$, and the appropriate string scale and infrared
cut-off:
\be
{16 \, \pi^2\over g^2_{\rm grav}(\mu^{(\rm As)})}=
- 2 \log \Im T_{\rm As} 
\left\vert \th_4 \left( T_{\rm As} \right)
\right\vert^4 -
2 \log \Im U_{\rm As} 
\left\vert \th_2 \left( U_{\rm As} \right)
\right\vert^4 + 4\log { M^{(\rm As)} \over \mu^{(\rm As)}}
\label{thrinta}
\ee
up to a constant\footnote{This constant, as well as the one appearing
later in the perturbative heterotic coupling (\ref{htr}), contains in
general $\log (M/\mu)$ terms, which account for extra massless states
that have been disregarded in the determination of the $R^2$ amplitude
from the $B_4$. Field-theoretical arguments can be advocated to fix these
terms. We will take care of them in the final expression (\ref{np}).}.  

This expression deserves some comments, as was for (\ref{thrint}).
We first observe that the $Z_2^{(1)}$-shift in
$\Gamma_{2,2}^{(1)}$ breaks the 
$SL(2,Z)_{T_{\rm As}} \times SL(2,Z)_{U_{\rm As}}\times 
Z_2^{T_{\rm As}\leftrightarrow U_{\rm As}}$ 
duality
group to a subgroup that depends on the kind
of shift performed. In the limit $\Im T_{\rm As} \to \infty$, $\Im
U_{\rm As}
 \to 0$
there is a restoration of an $N=(4,0)$ supersymmetry
with no linear behaviour either in $\Im  T_{\rm As}$
or in $1 \big/ \Im U_{\rm As}$; the remaining contribution is
logarithmic:
\be
{16 \, \pi^2\over g^2_{\rm grav}(\mu^{(\rm As)})}
\to  
- 2 \log \Im T_{\rm As}+ 2 \log \Im U_{\rm As}\, .
\label{logTa}
\ee
Finally, comparison of (\ref{thrint}) and  (\ref{thrinta})
suggests that the string duality map implies the following identification
of the moduli:
\be
T_{\rm As}\leftrightarrow T^2\sp U_{\rm As}\leftrightarrow
T^3 
{\rm \ and \ \ }4 \pi S_{\rm As}\leftrightarrow -1/T^1\, .
\label{symas}
\ee
The identification of the asymmetric dilaton $S_{\rm As}$ with the
$h^{1,1}$ moduli
$ T^1$ of the symmetric type II construction follows from the
behaviour in the
limit $\Im T^1\to 0$, which corresponds,
in the asymmetric construction,
to the perturbative limit
$S_{\rm As}\to \infty$.

\boldmath
\section{The $N=(2,0)$ heterotic construction}
\unboldmath

\subsection{\sl The construction of the model}

The heterotic dual to the previous type II constructions
is based on  $\left(T^2 \times T^4\right)\big/ Z_2^{(\rm f)}$ freely
acting orbifold heterotic
compactification. In order to reduce the gauge group we have to
introduce a set of ``discrete Wilson lines'',
which separate the boundary conditions of  the 32
right-moving  fermions, $\Psi_A,A=1,\ldots,32$, in order to twist all the
currents $\Psi_A \Psi_B$ of  the $c=(0,16)$ conformal block. The resulting
characters are described by those of 32 right-moving Ising's
. The $Z_2^{(\rm f)}$ projection
reduces the initial  $N=4$ supersymmetry to $N=2$; it acts
as a $\pi$ rotation in $T^4$ left  and right (super)coordinates and
as a translation in $T^2$. Its action on $\Psi_A$ is non-trivial
(see below), and it is chosen in such a way that neither vectors
nor hypers are produced from the $\Psi_A$'s. In this way the
heterotic massless spectrum is
identical to that of the previous type II constructions. Namely,
it consists of the $N=2$ supergravity multiplet,
1 vector-tensor multiplet that contains the dilaton $S_{\rm Het}$ 
and is dual to a vector, 2 vector multiplets 
associated with the K\"{a}hler class $T$ and the complex structure
$U$ of the torus\footnote{For simplicity we will use systematically 
$(T,U)$ for the heterotic two-torus moduli, instead of the more
natural notation, which would have been 
$(T_{\rm  Het},U_{\rm  Het})$.}
$T^2$, and 4 hypermultiplets,
obtained by pairing the left-moving negative eigenvalues of the
projection $Z_2^{(\rm f)}$ with the 4 right-moving negative
eigenvalues in $T^4$.

The partition function of the heterotic construction has the
following expression:
\ba
Z_{\rm Het} & = &
{1 \over \Im \tau   | \eta|^4 } {1 \over 2}
\sum_{H^{\rm f},G^{\rm f}}
Z_{6,22} \ar{H^{\rm f}}{G^{\rm f}} \nonumber \\
&& \times {1 \over 2} \sum_{a,b}{1\over \eta^4} (-)^{a+b+ab}
\vartheta \ar{a}{b}^2
\vartheta \ar{a+H^{\rm f}}{b+G^{\rm f}}
\vartheta \ar{a-H^{\rm f}}{b-G^{\rm f}} \, ,
\label{hH}
\ea
where the second line stands for
the contribution of the 10 left-moving
world-sheet fermions $\psi^{\mu},\Psi^I$ and the ghosts
$\beta,\gamma$ of the super-reparametrization;
$Z_{6,22}\ar{H^{\rm f}}{G^{\rm f}}$ accounts for the $(6,6)$
compactified coordinates and the $c=(0,16 )$ conformal system, which
is described by the
32 right-moving fermions $\Psi_A$. It
takes the
following form:
\be
Z_{6,22}\ar{H^{\rm f}}{G^{\rm f}}=
{1\over 2^{5}}\, \sum_{\vec h, \vec g}{1\over \eta^6 {\bar \eta}^6 }
\Gamma_{2,2} \ar{H^{\rm f},\vec h}{G^{\rm f},\vec g}
\,
\Gamma_{4,4} \ar{H^{\rm f}\vert \vec h}{G^{\rm f}\vert \vec g}\,
{\overline \Phi}\ar{H^{\rm f}, \vec h}{G^{\rm f}, \vec g}\, ,
\label{hH622}
\ee
where $(\vec h, \vec g)$ denote the five projections that are needed
in order to separate the boundary conditions of all 32 fermions.
The function $\Phi \ar{H^{\rm f}, \vec h}{G^{\rm f}, \vec g}$
can be written explicitly using the $SO(4)$ twisted characters
(see Ref. \cite{gkp}): 
\be
{\widehat{F}}_{1}\ar{\g,h}{\d, g} \equiv {1\over {\eta}^2}\,
{\vartheta}^{1/2}
\ar{\g-h_1-h_2-h_3}{\d-g_1-g_2-g_3}\,
{\vartheta}^{1/2}
\ar{\g+h_3}{\d+g_3}\,
{\vartheta}^{1/2}
\ar{\g+h_3-h_1}{\d+g_3-g_1}\,
{\vartheta}^{1/2}
\ar{\g+h_2-h_3}{\d+g_2-g_3}
\ee
and
\be
{\widehat{F}}_{2}\ar{\g, h}{\d, g} \equiv {1\over {\eta}^2}\,
{\vartheta}^{1/2}
\ar{\g}{\d}\,
{\vartheta}^{1/2}
\ar{\g+h_1-h_2}{\d+g_1-g_2}\,
{\vartheta}^{1/2}
\ar{\g+h_1}{\d+g_1}\,
{\vartheta}^{1/2}
\ar{\g+h_2}{\d+g_2}\, ,
\ee
where we introduced the notation
$h\equiv (h_1,h_2,h_3)$ and similarly for $g$.
Under $\tau\to \tau+1$, ${\widehat{F}}_{I}$
transform as:
\ba
{\widehat{F}}_{1}\ar{\g, h}{\d, g} & \to &
{\widehat{F}}_{1}\ar{\g, h}{\g+\d +1, h + g} \nn \\
&& \times
\exp -{i\pi\over 4}
\left({2\over 3} +2 \gamma^2+h_1^2+h_2^2 +2 h_3^2 - 2 \g h_1 + h_1 h_2
- 4\g +2 h_1 \right) \, ;
\\
{\widehat{F}}_{2}\ar{\g, h}{\d, g} & \to &
{\widehat{F}}_{2}\ar{\g, h}{\g+\d +1, h + g} \nn \\
&& \times
\exp -{i\pi\over 4}
\left({2\over 3} +2 \gamma^2+h_1^2+h_2^2  + 2 \g h_1 - h_1 h_2
- 4 \g  -2 h_1 \right)\, .
\ea
Notice that ${\widehat{F}}_{I}$ are $c=(2,0)$ conformal characters of 4
different left-moving Isings. In the fermionic language \cite{abk}
this is a system of 4 left-moving real fermions with different
boundary
conditions.
All currents ${J}^{IJ}={ \Psi}^I{ \Psi}^J$ are twisted
and therefore the initial $SO(4)$ is broken.
We have then, as in \cite{gkp}, two alternative constructions,
$\Phi$ and $\tilde{\Phi}$, that differ with respect to the embedding of
the
$Z_2^{(\rm f)}$-shift in the two-torus:
\ba
\Phi \ar{H^{\rm f}, \vec h}{G^{\rm f}, \vec g} & = &
{1 \over 2} \sum_{\g,\d}\,
{\widehat{F}}_{1} \ar{\g+H^{\rm f}, h}{\d+G^{\rm f}, g} \,
{\widehat{F}}_{1} \ar{\g+h_4+H^{\rm f}, h}{\d+g_4+G^{\rm f}, g} \,
{\widehat{F}}_{1} \ar{\g+h_5, h}{\d+g_5, g} \,
{\widehat{F}}_{1} \ar{\g+h_4+h_5, h}{\d+g_4+g_5, g} \nn \\
&& \times \, {\widehat{F}}_{2} \ar{\g, h}{\d, g} \,
{\widehat{F}}_{2} \ar{\g+h_4, h}{\d+g_4, g} \,
{\widehat{F}}_{2} \ar{\g+h_5, h}{\d+g_5, g} \,
{\widehat{F}}_{2} \ar{\g+h_4+h_5, h}{\d+g_4+g_5, g}
\ea
and
\ba
\tilde{\Phi} \ar{H^{\rm f}, \vec h}{G^{\rm f}, \vec g} & = &
{1 \over 2} \sum_{\g,\d}\,
{\widehat{F}}_{1} \ar{\g+H^{\rm f}, h}{\d+G^{\rm f}, g} \,
{\widehat{F}}_{1} \ar{\g+h_4+H^{\rm f}, h}{\d+g_4+G^{\rm f}, g} \,
{\widehat{F}}_{1} \ar{\g+h_5, h}{\d+g_5, g} \,
{\widehat{F}}_{1} \ar{\g+h_4+h_5, h}{\d+g_4+g_5, g} \nn \\
&& \times \,
{\widehat{F}}_{2} \ar{\g+H^{\rm f}, h}{\d+G^{\rm f}, g} \,
{\widehat{F}}_{2} \ar{\g+h_4, h}{\d+g_4, g} \,
{\widehat{F}}_{2} \ar{\g+h_5, h}{\d+g_5, g} \,
{\widehat{F}}_{2} \ar{\g+h_4+h_5, h}{\d+g_4+g_5, g} \, .
\label{88fivt}
\ea

The $(2,2)$ and $(4,4)$ lattice shifts are dictated by modular
invariance and are needed in order
to cancel the phases that appear under
modular transformations. These shifts are different for the two
constructions,
based on $\Phi$ or $\tilde{\Phi}$.  In the case of $\Phi$, modular
invariance implies an asymmetric shift on the
$\Gamma_{2,2} \ar{H^{\rm f}}{G^{\rm f}}$, which we chose to be
$(-)^{(m_1+n_1) G^{\rm f}}$ (this projection was referred to as ``X''
in Ref. \cite{kkprn}, where the various lattice shifts were discussed
in detail); the
shift in $\Gamma_{4,4}\ar{H^{\rm f}\vert \vec h}{G^{\rm f}\vert \vec
g}$, however, has to be symmetric and is chosen to be $(-)^{M_i \,
g_i}$.

In the construction based on $\tilde{\Phi}$, on the other hand,
the $(2,2)$ lattice must be doubly shifted, $\Gamma_{2,2}\ar{H^{\rm
f}, h_1-h_2}{G^{\rm f}, g_1-g_2}$.
In this case we use the projection $(-)^{m_1G^{\rm f}+n_1(g_1-g_2)}$.

Constructions $\Phi$ and $\tilde{\Phi}$
share the same $N=4$ sector (defined by $(H^{\rm f},G^{\rm f})=(0,0)$).
The contribution of this sector to the partition function is one half
of the partition function of an $N=4$ model
in which the gauge group is $U(1)^6$, and all the vectors originating
from the torus $T^6$. In the  $N=2$ sectors,
$(H^{\rm f},G^{\rm f})\neq (0,0)$, while $(h_i, g_i)$ are either
$(0,0)$ or $(H^{\rm f},G^{\rm f})$. This restriction on the values
$(h_i, g_i)$ comes from the
$(H^{\rm f},G^{\rm f})$-twisted
sector of $\Gamma_{4,4} \ar{H^{\rm f}\vert \vec h}{G^{\rm f}\vert
\vec g}$.

A  quantity relevant to our purpose is the helicity
supertrace $B_2$, which receives a non-zero contribution from the $N=2$
sector, while the contribution  of the $N=4$ sector to this
quantity vanishes.
For the construction based on $\Phi$ we find:
\be
B_2 \left(\Phi \right)={1\over \bar{\eta}^{24}}\,
\sumpf
\Gamma_{2,2}^{\lambda=1} \ar{H^{\rm f}}{G^{\rm f}} \,
\bOmega \ar{H^{\rm f}}{G^{\rm f}}\, ,
\label{B288}
\ee
where $\Omega \ar{H^{\rm f}}{G^{\rm f}}$ turn out to be
the same analytic functions as for the models
considered in Ref.~\cite{gkp}\footnote{The parameter $\l$, which
takes the
values 0 or 1, determines the phases appearing in the modular
transformations of the shifted lattice sums. 
These phases are complementary of those coming
from the corresponding functions $\Omega\ar{H^{\rm f}}{G^{\rm f}}$ or  
$\Omega^{(0)}\ar{H^{\rm f}}{G^{\rm f}}$ and 
$\Omega^{(1)}\ar{H^{\rm f}}{G^{\rm f}}$.}, even though
the $N=2$ constructions are 
different\footnote{This is not surprising since the $N=2$ models under
consideration are known to fall in a very restricted set of universality 
classes with respect to their elliptic genus \cite{kkprn}.}: 
\ba
\Omega \ar{0}{1}&=&\hphantom{-}{1\over 16}
\left({\vartheta}_3^8 + {\vartheta}_4^8
+14\, {\vartheta}_3^4\, {\vartheta}_4^4 \right)
{\vartheta}_3^6\,  {\vartheta}_4^6\nn \\
\Omega \ar{1}{0}&=&-{1\over 16}
\left({\vartheta}_2^8 + {\vartheta}_3^8
+14\, {\vartheta}_2^4\, {\vartheta}_3^4 \right)
{\vartheta}_2^6\,  {\vartheta}_3^6\label{Om88} \\
\Omega \ar{1}{1}&=&\hphantom{-}{1\over 16}
\left({\vartheta}_2^8 + {\vartheta}_4^8
-14\, {\vartheta}_2^4\, {\vartheta}_4^4 \right)
{\vartheta}_2^6\,  {\vartheta}_4 ^6\, .\nn
\ea

The construction based on $\tilde{\Phi}$ with $N_V=N_H=0$, 
under study here, has the
same universality properties as the corresponding $N=2$ models of 
Ref. \cite{gkp}. The
helicity supertrace $B_2$ is identical for all such models with $N_V=N_H$,
irrespectively of whether the latter 
vanishes or not:
\be
B_2 \left(\tilde{\Phi} \right)=
{1\over \bar{\eta}^{24}}\,
\sumpf {1\over 2}\left(
\Gamma_{2,2}^{\lambda=0} \ar{H^{\rm f}}{G^{\rm f}} \,
\bOmega^{(0)} \ar{H^{\rm f}}{G^{\rm f}} +
\Gamma_{2,2}^{\lambda=1} \ar{H^{\rm f}}{G^{\rm f}} \,
\bOmega^{(1)} \ar{H^{\rm f}}{G^{\rm f}}
\right) \, ,
\label{B288t}
\ee
where in this case
\ba
\Omega^{(0)}\ar{0}{1}&=&{\hphantom{-}}
\frac{1}{2}
\left(\th_3^4+\th_4^4\right)
\th_3^8\, \th_4^8
\nonumber\\
\Omega^{(0)}\ar{1}{0}&=&-
\frac{1}{2}
\left(\th_2^4+\th_3^4\right)
\th_2^8\, \th_3^8
\label{Om88a}\\
\Omega^{(0)}\ar{1}{1}&=&{\hphantom{-}}
\frac{1}{2}
\left(\th_2^4-\th_4^4\right)
\th_2^8\, \th_4^8\nonumber
\ea
and
\be
\Omega^{(1)} \ar{H^{\rm f}}{G^{\rm f}}
= (-)^{H^{\rm f}}\left(
\omega \ar{H^{\rm f}}{G^{\rm f}}\right)^{10} \, ,
\ee
with
\be
\omega \ar{0}{1}= \th_3 \, \th_4 \sp
\omega \ar{1}{0}= \th_2 \, \th_3 \sp
\omega \ar{1}{1}= \th_2 \, \th_4 \, .
\label{Om88b}
\ee
The lattice sums $\Gamma_{2,2}^{\lambda=0} \ar{H^{\rm f}}{G^{\rm f}}$
and $\Gamma_{2,2}^{\lambda=1} \ar{H^{\rm f}}{G^{\rm f}}$
correspond to simply shifted lattices with projections
$(-)^{m_1 G^{\rm f}}$ and $(-)^{(m_1+n_1) G^{\rm f}}$, respectively (the
cases ``I'' and ``X'' of \cite{kkprn}).

In both constructions  $\Phi$ and $\tilde \Phi$, the massless
contribution to the $B_2$ vanishes for generic values of the moduli
$T$ and $U$,  as it should for models where
$N_V = N_H$. Owing to the
$Z_2^{(\rm f)}$-translation on the two-torus,
the $N=4$ supersymmetry is spontaneously broken.
The analysis of these constructions is the same as for the
models discussed in \cite{gkp}, to which we refer for the details.
Here we simply recall that the construction based on $\Phi$
is not suitable for a comparison with the type II ground states:
the region in the space of (discrete) Wilson lines which allows for an
easy identification of the map
between the moduli of the dual models is the one that corresponds to
the construction based on $\tilde{\Phi}$.
In the latter, and for
the particular $Z_2^{(\rm f)}$-shift we have considered,
$(-)^{m_1 G^{\rm f}+n_1 g_1}$, the mass of the two extra gravitinos is
\be
m^2_{3/2} = {1 \over 4} {  \Im U \over \Im T}\, .
\label{mhet}
\ee
The $N=4$ supersymmetry is restored
when  $R_1= \sqrt{ \Im T / \Im U}$ is large.
For large values
of $\Im U / \Im T$ we recover instead a genuine $N=2$
non-freely-acting orbifold.
For the specific directions of the shifts in the
two-torus that we have chosen, there are lines in the $(T,U)$-plane
along which two extra hypermultiplets appear in the massless spectrum
together with two extra massless vectors leading to
an $SU(2)$ enhancement of one of the $U(1)$'s of the torus.

\subsection{\sl The gravitational corrections}

In order to obtain heterotic gravitational
corrections analogous to those of the type II constructions (see Eqs.
(\ref{thrint}) and (\ref{thrinta})), we must
proceed as follows. Instead of considering the
pure $R^2$ term, we must compute the one-loop corrections for a
special combination of gravitational and helicity operators.
The ordinary $R^2$-term correction is given by the genus-one 
amplitude of \be
Q_{\rm grav}^2\equiv  Q^2\,
\bP_{\rm grav}^2\, , 
\label{Qgrav}
\ee
where $Q$ stands again for the left-helicity operator, and 
$\bP_{\rm grav}^2$ is the usual gravitational operator: 
when inserted in the one-loop vacuum amplitude, it acts as 
${-1\over 2 \pi i}{\partial \over
\partial{\bar \tau}}$ on ${1/ \im \, \bar{\eta}^2}$;
namely, it acts on the contribution of the two right-moving transverse
space-time coordinates $\bar X^{\mu=3,4}$, {\it including their
zero-modes.} This latter fact is responsible for the appearance of 
a non-holomorphic gravity-backreaction
contribution, which ensures modular covariance but has no type II
counterpart, as was discussed in \cite {gkp}; the one-loop
amplitude of the above operator is\footnote{In general the
heterotic one-loop amplitude of an operator of the form $Q^2\, \bP^2$
reads:
$
\left\langle Q^2\, \bP^2
\right\rangle_{\rm genus-one}
= \bP^2\,B_2\, ,
$
where, in the l.h.s., $\bP^2$ acts as a differential operator on some
specific factor of $B_2$.}
\be
F_{\rm grav}\equiv \left\langle Q_{\rm grav}^2 
\right\rangle_{\rm genus-one}
= -{1\over 12}\,  
\left( \bE_2 -{3\over \pi\im}\right) B_2\, ,
\label{Fgrav}
\ee
where, for the model under consideration (constructed with
$\tilde \Phi$), $B_2$ is given in Eq. (\ref{B288t}). The massless
contribution to $F_{\rm grav}$ is precisely the gravitational anomaly,
$b_{\rm grav}= {24 - N_V +N_H\over 12}$, which in the case
at hand equals 2 ($B_2\vert_{\rm massless}$ vanishes), at generic
points of the $(T,U)$ moduli space.

The operator $\bP_{\rm grav}^2$ is not suitable for comparison with 
the type II result, because it is not holomorphic and its amplitude is
sensitive to the
$N=2$ singularities occurring in the $(T,U)$ plane: the corresponding
beta-function
(i.e. the gravitational
anomaly) jumps along rational lines where 
$\Delta N_V\neq\Delta N_H$.  We must therefore replace
$\bP_{\rm grav}^2$ with an appropriate holomorphic operator 
$\bP_{\rm grav}^{\prime 2}$ whose amplitude 
is regular everywhere in $(T,U)$, at least in the model constructed
with 
$\tilde \Phi$, which is the model we will be analysing in the
following.
To this purpose, we introduce two
operators:
$\bHt$ and $\bP^2_{2,2}$.

The operator $\bHt$ acts, for any $(H^{\rm f},G^{\rm f})$-twisted
sector of the orbifold, as a derivative
${ -1 \over 2 \pi i} {\partial \over \partial \bar{\tau}}$
on the factor $\bar\omega \ar{H^{\rm f}}{G^{\rm f}}
\big/ \Im \tau \, \bar{\eta}^4$, which contains the contribution of
twisted coordinates (see Eq. (\ref{Om88b})). After some 
straightforward algebra we obtain the amplitude:
\be
\bHt\, B_2 \left(\tilde{\Phi} \right)=
-{1\over 24}\sum_{\l=0,1}\sumpf
\Gamma_{2,2}^{\lambda} \ar{H^{\rm f}}{G^{\rm f}} 
\left(
\bE_2 -{3\over \pi\im}
+{1\over 2}\bH\ar{H^{\rm f}}{G^{\rm f}}
\right)
{\bOmega^{(\l)} \ar{H^{\rm f}}{G^{\rm f}}\over
\bar{\eta}^{24}}\, ,
\label{bHt}
\ee
where we have introduced the modular-covariant functions
\be
H{h\atopwithdelims[]g}={12 \over  \pi i} \partial_\t \log {\th
{1-h\atopwithdelims[]1-g}\over
\eta}
= \cases{\hphantom{-} \th_3^4 + \th_4^4  \sp (h,g)=(0,1)\cr
          -  \th_2^4 - \th_3^4  \sp (h,g)=(1,0)\cr
\hphantom{-} \th_2^4 - \th_4^4  \sp (h,g)=(1,1)\cr }
\label{H}
\ee
of weight $2$. In the model constructed with ${\Phi}$, only a 
$\lambda =1$ term would
appear in (\ref{bHt}).

From expression (\ref{bHt}) we observe that the insertion of 
$\bHt$ is covariant but not holomorphic. This latter property 
will allow for cancelling the non-holomorphic terms present when 
$\bP_{\rm grav}^2$ and $\bP^2_{2,2}$ are inserted in the vacuum
amplitude, while keeping modular covariance.
The beta-function
coefficient of this operator, i.e. the constant term of $\bHt\, B_2
\left(\tilde{\Phi}
\right)$, vanishes for  generic $(T,U)$ while it jumps accross several
special lines:
$\Delta b\left(\bHt\right) = 2\, \Delta b_{\rm grav}$. 

On the other hand, after insertion into the one-loop heterotic vacuum
amplitude,
$\bP^2_{2,2}$ acts as ${-1 \over 2 \pi i}{\partial \over
\partial{\bar \tau}}$ on the modular-covariant factor of weight zero,
$\im \,\Gamma_{2,2}^{\lambda}\ar{H^{\rm f}}{G^{\rm f}}$.
This amounts to inserting the sum of the two right-moving
lattice momenta ${\bar p}_1^2+{\bar p}_2^2$ of $T^2$, which correspond
to the Cartan of the $U(1)$ factor. The amplitude 
$\left\langle Q^2\, \bP^2_{2,2}
\right\rangle$ therefore generates the corresponding gauge-coupling
correction. To be more precise, we must in fact consider the integral  
$\int_{\cal F} {d^2 \tau \over \im}\left\langle Q^2\, \bP^2_{2,2}
\right\rangle \g (\tau,\bar \tau)$, where $\g (\tau,\bar \tau)$ is an
appropriate modular-invariant infrared-regularizing function
\cite{infra}. An integration by parts can be performed, which leads
to vanishing boundary terms {\it all over the $(T,U)$ plane},
irrespectively of the specific behaviour of the lattice sums across
rational lines. This allows us to recast the above amplitude as:
\be
\bP^2_{2,2}\, B_2 \left(\tilde{\Phi} \right)=
{1\over 2}
\sum_{\lambda=0,1}\sumpf
\left(
\Gamma_{2,2}^{\lambda} \ar{H^{\rm f}}{G^{\rm f}} 
\left[
{1 \over 2 \pi i}{\partial \over
\partial{\bar \tau}}- {1\over 2\pi \im}
\right]
{\bOmega^{(\lambda)} \ar{H^{\rm f}}{G^{\rm f}}\over \bar{\eta}^{24}}
\right) 
\label{P22}
\ee
(in the model constructed with ${\Phi}$, only a $\lambda =1$ term
appears). The differential operator inside the brackets is covariant, since
${\bOmega^{(\lambda)} \ar{H^{\rm f}}{G^{\rm f}}\Big/ \bar{\eta}^{24}}$
has modular weight $-2$, but non-holomorphic. The beta-function
coefficient $b(P_{2,2})$ (constant term in (\ref{P22})) vanishes for
generic $(T,U)$  and its discontinuity at special lines turns out to
be $\Delta b(P_{2,2})= -12\, \Delta b_{\rm grav}$.

Given the operators $\bP_{\rm grav}^2$, $\bHt$ and $\bP^2_{2,2}$,
there is a {\it unique} combination, which is holomorphic and whose
beta-function coefficient is equal to the gravitational anomaly
$b_{\rm grav}=2$ {\it everywhere} in the moduli space $(T,U)$:
\be
\bP_{\rm grav}^{\prime 2}=\bP_{\rm grav}^2
-{5\over 4}
\bHt
-{1\over 8}
\bP^2_{2,2}
\, .\label{reg}
\ee
Here we want to remark that,
in some specific cases where 
$N_V=N_H \neq 0$,
as for the models considered
in \cite{gkp}, the operator $\bHt$
can be reexpressed through an integration
by parts, as a combination of
$\overline{P}^2_{2,2}$ and
$\overline{P}^2_{\rm gauge}$, the gauge operator for the
higher-level currents (here absent).
In such cases, expression (\ref{reg}) becomes 
$\overline{P}^2_{\rm grav}+{1 \over 12}\overline{P}^2_{2,2}
+{5 \over 3 N_V} \overline{P}^2_{\rm gauge}$.

The amplitude $ \left\langle Q_{\rm grav}^{\prime 2}\right\rangle=
\left\langle Q^2\, \bP_{\rm grav}^{\prime 2}
\right\rangle $ at genus one reads:
\be
\bP_{\rm grav}^{\prime 2}\,B_2\left(\tilde \Phi\right)=2\sumpf
\Gamma_{2,2}^{\lambda=0} \ar{H^{\rm f}}{G^{\rm f}} 
\, ,
\ee
which is regular everywhere. Notice that the contribution of the
$\lambda=1$ term vanishes identically. This amplitude leads to the
thresholds
\be
\Delta_{\rm Het}=2 \, \int_{\cal F} {d^2 \tau \over \Im \tau}
\left( \sumpf
\Gamma_{2,2}^{\lambda=0} \ar{H^{\rm f}}{G^{\rm f}}-1 \right)\, , 
\ee
which is valid only for the construction based on $\tilde \Phi$.

For the specific case in which the shift 
in $\Gamma_{2,2}^{\l = 0}$ is due to a translation of momenta,
$(-1)^{m_1 G^{\rm f}}$, we get:
\be
\Delta_{\rm Het}= -2  \log \Im T \left\vert \th_4 \left( T \right)
\right\vert^4
- 2  \log \Im U \left \vert \th_2 \left( U \right)
\right\vert^4 +  {\rm const.}
\label{dhet}
\ee
Finally the running of the coupling is given by
\ba
{16 \, \pi^2 \over g^2_{\rm grav}\left( \mu^{(\rm Het)} \right)}
& = & 16 \, \pi^2  \kappa  \Im S_{\rm Het}  
-2 \log \Im T \left\vert \th_4 \left( T \right) \right\vert^4 
- 2 \log \Im U \left \vert \th_2 \left( U \right) \right\vert^4
\nn \\
&& +4 \log {M^{(\rm Het)} \over \mu^{(\rm Het)} }
 + {\rm const.}\, ,
\label{htr}
\ea
where $S_{\rm Het}$ is the heterotic dilaton--axion field and
\be
\kappa \, \Im S_{\rm Het} ={1\over g^2_{\rm Het} }\, .
\label{dilk}
\ee
In the limit where $(1/ \Im T, \Im U)\to 0$, $N=4$ supersymmetry is
restored with the following behaviour:
\be
{16 \, \pi^2 \over g^2_{\rm grav}\left( \mu^{(\rm Het)} \right)}
=16 \, \pi^2 \kappa \Im S_{\rm Het}  
-2 \log \Im T +2 \log \Im U 
\, .
\label{htrli}
\ee
The role of the  normalization factor $\kappa$ will be
specified in the next section
where  the gravitational corrections of the heterotic $\tilde{\Phi}$
construction and that of the two
type II ground states will be compared.

\section{Comparison of the three duals}

In this section we would like to test the triality relation between 
all
three freely acting orbifolds we have considered, through the
analysis of the
``gravitational'' corrections.
First of all we observe that the operator
$Q_{\rm grav}^{\prime 2}= Q^2\, \bP_{\rm grav}^{\prime 2}$
with $\bP_{\rm grav}^{\prime 2}$
given in (\ref{reg}), coincides, on type II side, with the
operator $Q^{{\rm II}2}_{\rm grav}= 2 Q^2 \overline{Q}^2$ we considered
in Sections 2 and 3. Indeed, owing to the absence of
perturbative Ramond--Ramond charges, the contribution of the dual of
$\overline{P}^2_{\rm grav}$ vanishes; because of the symmetry
between left- and right-movers on the world-sheet, there is no
need for us to introduce an operator such as $\bHt$,  the insertion of
$Q^{{\rm II}2}_{\rm grav}$ being automatically holomorphic.
The duality among the three orbifolds requires
the identification of
one of the three perturbative
vector multiplet moduli of the type IIA symmetric orbifold, with the
dual of the field $S_{\rm Het}$, the dilaton--axion field
of the heterotic theory, and with the inverse of
$S_{\rm As}$, the dilaton--axion field of the type II
asymmetric theory.
Modulo $SL(2,Z)$ transformations,
such modulus can be indifferently any one of
the three $T^i$, $i=1,2,3$. For definiteness, we will choose $T^1$,
as was anticipated in (\ref{symas}). In order to see what the
precise duality map is, we consider all three models in their
``$N=4$ phase''. In this limit, the heterotic amplitude 
$\left\langle Q_{\rm grav}^{\prime 2}\right\rangle$ is expected
to receive contributions from genus zero only, while in the type II
$\left\langle Q_{\rm grav}^{{\rm II} 2}\right\rangle$
should vanish in the asymmetric orbifold and depend on one complex
modulus only in the type IIA symmetric one.
This behaviour can be checked by taking the appropriate
$N=4$ limits  in the three models
(see Eqs. (\ref{logT}), (\ref{linear}), (\ref{logTa}) and
(\ref{mhet}), (\ref{htr})).
The other surviving contributions, with a logarithmic dependence
on the other moduli, are in fact
infrared artefacts due to an accumulation of massless
states, which can be lifted, in all the three models,
by switching on
an infrared cut-off larger than the massive gravitinos,
as we previously discussed.
By comparing Eq. (\ref{linear}) with the genus-zero contribution in
Eq. (\ref{htrli})  ($16 \, \pi^2 \kappa \, \Im S_{\rm Het} $), we obtain
\be
T^1={\kappa \over 2} \tau_{S_{\rm Het}}=
2 \pi  \kappa S_{\rm Het} \, .
\ee
Here we have introduced the field $\tau_S \equiv 4 \pi S$,
the actual ``modular'' parameter of the Montonen--Olive
$S$-duality transformations. Since, after an $SL(2,Z)_T$
inversion $T \to -1/T$,  the model becomes symmetric under permutations
of the fields
$\tau_{S_{\rm Het}}, T$ and $U$, the behaviour of the effective
coupling constant, for large values of the three moduli, must be
symmetric as well; this requirement forces us to fix $\kappa=2$.
This normalization of the coupling (\ref{dilk}), which differs by a
factor 2 from the usual tree-level coupling (corresponding to
$\kappa=1$), is required also for a correct interpretation in terms
of instanton contributions (see below).
In the opposite limit, $T^1 \to 0$,
the $T^1$-dependent contribution vanishes (up to an irrelevant
logarithmic term in $-1 / T^1$).
This is consistent with the identification of $-1/T^1$
with $\tau_{S_{\rm As}}= 4 \pi S_{\rm As}$, and 
implies in particular that {\it the type II asymmetric orbifold
is the strong coupling limit of the heterotic}:
\be
\tau_{S_{\rm As}}= -{1 \over \tau_{S_{\rm Het}}}\, .
\ee

In order to test the above duality relations, we consider the
``$N=2$ phase'' of the various theories under consideration, where 
the dependence on the other moduli, generating from genus one for all
three models, remains.
The part of the gravitational amplitude
that depends on the perturbative moduli is indeed the same
in the three models, provided we identify the moduli $(T_{\rm
As},U_{\rm As})$ with $(T^2,T^3)$ and $(T,U)$.
Through the duality map, we therefore learn that,
similarly to the symmetric type IIA orbifold, the
heterotic model possesses an $N=8$ supersymmetry that is
broken spontaneously.
This breaking on the heterotic side is non-perturbative: the Higgs
field whose vev is the order parameter for the spontaneous breaking of
the $N=8$ supersymmetry to $N=4$
is the dilaton $S_{\rm Het}$, and the $N=8$ supersymmetry can be
restored 
at the strong coupling limit.
The further breaking of the supersymmetry from $N=4$ to $N=2$
is realized spontaneously in a perturbative way, by using as
Higgs fields the moduli $T$ and $U$.
The full, perturbative and non-perturbative, correction to
the effective coupling constant of the gravitational term
considered here
is given by the type IIA result, Eq. (\ref{thrint}).
In heterotic string variables, we have:
\ba
{16 \, \pi^2 \over g^2_{\rm grav} \left( \mu \right)}
& = & -2 \log \Im \tau_{S_{\rm Het}} \left\vert \th_2
\left(  \tau_{S_{\rm Het}} \right) \right\vert^4 \nn \\
&& -2 \log \Im T \left\vert \th_4 \left( T \right)
\right\vert^4
-2 \log \Im U\left\vert
\th_2 \left( U\right)\right\vert^4 \nn \\
&& + 6 \log { M_{\rm Planck} \over \mu} + {\rm const.} \, ,
\label{np}
\ea
where we have expressed the infrared running in terms
of the duality-invariant Planck mass and the physical cut-off
$\mu$, related to the various string scales by:
\be
{M_{\rm Planck} \over \mu}={M^{(\rm Het)} \over \mu^{(\rm Het)}}=
{M^{(\rm IIA)} \over \mu^{(\rm IIA)}}=
{M^{(\rm As)} \over \mu^{(\rm As)}}\, .
\ee
From expression (\ref{np}) we can easily read off the instanton
numbers $k$,
given by the powers of $q \equiv \exp 2 \pi i \tau_{S_{\rm Het}}$ in the
expansion of the
first term. We obtain $k \in {\kappa N \over 2}$, 
which for $\kappa=2$ becomes, as expected, $k \in N$.

\section{The prepotential}

\subsection{\sl The one-loop result}

The perturbative prepotential can be easily computed from the
heterotic side. Owing to the $N=2$ supersymmetry, it receives no
perturbative corrections beyond one loop.
The tree-level contribution is determined by the geometric properties
of the vector manifold, and is the same for both the constructions
based on $\Phi$ and on $\tilde{\Phi}$ \cite{wkll}: 
\be
h^{(0)}=-{ i \over 2 \pi} \tau_{S_{\rm Het}} T  U\, .
\label{h0}
\ee
The genus-one correction, on the other hand,
is different in the two constructions and, moreover, it
depends on the choice of shift vectors in the two-torus.
Here, we will concentrate on the choices made previously
for these shift vectors. The
one-loop corrections to the prepotential turn out to satisfy
second-order differential equations.
These equations are obtained by properly treating the universal part
of the gauge corrections in models with spontaneously broken
supersymmetries, thereby generalizing the approach of
\cite{delgrav, hmbps, agn}.
For the models based on $\Phi$ the correction
$h^{(1)}$  solves
\ba
\lefteqn{{\rm Re} \Bigg( -{1 \over 2 T_2 U_2}
\left( 1-iU_2 {\partial \over \partial U} \right)
%&\! \!   &
\left( 1-iT_2 {\partial \over \partial T} \right)
h^{(1)} \Bigg)}
\nn \\
&&~~~~~~~~~~~~~~~~~~~~~~~~~~~~ =
{1 \over 64 \pi^3} \int_{\cal F} {d^2 \tau \over \tau_2}
\sump \Gamma_{2,2}^{\l = 1} \ar{h}{g}
\left(  i {d \over d \bar{\tau}}  + {1 \over \tau_2} \right)
{ \overline{\Omega} \ar{h}{g} \over \bar{\eta}^{24} }\, ,
\label{A}
\ea
while for models based on $\tilde{\Phi}$ it solves
\ba
\lefteqn{{\rm Re} \Bigg( -{1 \over 2 T_2 U_2}
\left( 1-iU_2 {\partial \over \partial U} \right)
\left( 1-iT_2 {\partial \over \partial T} \right)
\tilde h^{(1)} \Bigg) }
\nn \\
&&~~~~~~~~~~~~~~~~~~~~~~~~~~~~ =
{1 \over 64 \pi^3} \int_{\cal F} {d^2 \tau \over \tau_2}
{1 \over 2} \sum_{\l=0,1 } \sump  \Gamma^{\l}_{2,2} \ar{h}{g}
\left(  i {d \over d \bar{\tau}}  + {1 \over \tau_2} \right)
{ \overline{\Omega}^{(\l)} \ar{h}{g} \over \bar{\eta}^{24} }\, .
\label{B}
\ea
Notice that,
in contrast to the universal corrections \cite{kkprn},
the r.h.s. of the above equations has singularities across lines
in the moduli space. Integrals of this kind and analysis of the
singularities have been performed in several papers.
We will not present the general result here but give instead the
answer for the
prepotential. It is important to observe that the above equations
define $h^{(1)}$ up to irrelevant linear and quadratic terms as well
as cubic terms such as $T^2 U$ or $U^2 T$. These ambiguities can be
resolved by looking at the ordinary gravitational threshold
corrections \cite{kkprn, hmbps}:
\be
\Delta_{\rm grav}
\left(\Phi\right)
= -{1 \over 12} \int_{\cal F} {d^2 \tau \over \tau_2}
\left(  \sump \Gamma_{2,2}^{\l = 1} \ar{h}{g} \widehat{E}_2
{\overline{\Omega} \ar{h}{g} \over \bar{\eta}^{24} }
+12 \, b_{\rm grav}   \right)
\label{C}
\ee
or
\be
\Delta_{\rm grav}
\left(\tilde{\Phi}\right)
= -{1 \over 12} \int_{\cal F} {d^2 \tau \over \tau_2}
\left(  {1 \over 2} \sum_{\l=0,1}
\sump \Gamma_{2,2}^{\l} \ar{h}{g} \widehat{E}_2
{\overline{\Omega}^{(\l)} \ar{h}{g} \over \bar{\eta}^{24} }
+12 \, b_{\rm grav}   \right)\, .
\label{D}
\ee

We would like to stress at this point that Eqs. (\ref{A}) and (\ref{C})
(resp. (\ref{B}) and (\ref{D})) hold for heterotic constructions of the
kind  $\Phi$ (resp.  $\tilde{\Phi}$) with $N_V=N_H\neq 0$, as those
presented in \cite{gkp}. Therefore, our results for the perturbative
prepotential  $h^{(1)}$ or  $\tilde h^{(1)}$ given below (Eqs. (\ref{h})
and (\ref{htilde})) are valid for these more general models. This is again
due to the fact that the heterotic ground states under consideration fall
into the same elliptic-genus universality class, irrespectively of the
value of $N_V=N_H$. Non-perturbative contributions, however, depend on the
number of vector multiplets and hypermultiplets (through, for example, the
instanton numbers), and only the case
$N_V=N_H=0$ is analysed in the following.

After some lengthy algebra, we can solve the above equations and obtain
the one-loop prepotential
for models based on the construction $\Phi$:
\ba
h^{(1)}(T,U) & = & -{1\over (2 \pi )^4 }
\left(
{\cal L}_c(T,U)+{\cal L}_a(T,U)+{\cal L}_b(T,U)
\right)
- {i \over 8 \pi} T^2 U
\sp
{\rm for}
\ \
T_2 > U_2 \nn \\
& = &  - {1 \over (2 \pi )^4 }
\left(
{\cal L}_c(U,T)+{\cal L}_a(T,U)+{\cal L}_b(U,T)
\right)
- {i \over 8 \pi} T U^2
\sp
{\rm for}
\ \
T_2 < U_2\, . \nn \\
&& \label{h}
\ea
The functions ${\cal L}_{c,a,b}(T,U)$ are given in the appendix;
${\cal L}_{c,b}(T,U)$
have a branch along $T=U$, where
$\left. \Delta B_2\right\vert_{\rm massless}=
\Delta N_V - \Delta N_H= -14$. For models based on the construction
$\tilde \Phi$, we obtain:
\ba
\tilde{h}^{(1)}(T,U) & = & -{1 \over (2 \pi)^4} {1 \over 2}
\Bigg(-{\cal L}^{(0)}_c(T,U)
+{\cal L}^{(0)}_a\left({T\over 2},2U\right)
+{\cal L}^{(0)}_b\left({T\over 2},U\right)
\nn \\
&&\ \ \ \ \ \ \ \  \ \ \ +{\cal L}^{(1)}_c(T,U)
+{\cal L}^{(1)}_a(T,U) +{\cal L}^{(1)}_b(T,U)
\Bigg)
\, ,
\label{htilde}
\ea
where the functions ${\cal L}^{(\l)}_{c,a,b}(T,U)$
are as displayed in the appendix.
In this case there is no branch at $T=U$, where now $\left. \Delta
B_2\right\vert_{\rm massless}= 0$, and (\ref{htilde}) is thus valid
for any $T$ and $U$.  This makes the monodromy trivial around $T=U$.
Remember, however, that in the models at hand, where the two-torus
lattices are shifted, the target-space duality group is only a subgroup of 
$SL(2,Z)_T \times SL(2,Z)_U \times Z_2^{T\leftrightarrow U}$. In
particular $T\to -1/T$ is not a symmetry, and the line $T=U$ is not
equivalent to the line $-1/T=U$. The latter, where $\left. \Delta
B_2\right\vert_{\rm massless}= 2$, is a branch for $\tilde{h}^{(1)}(T,U)$,
although this is not straightforward from expression (\ref{htilde}) -- a
Poisson resummation is needed.
Observe also the absence of cubic terms in $\tilde{h}^{(1)}$. Such terms
are present in generic models with spontaneously broken
supersymmetry, as those studied in \cite{kkprn}, for which the
corrections to the prepotential can be computed in a similar way.
Cubic terms vanish in general when the intersection form of the
underlying Calabi--Yau manifold of the type II dual becomes
trivial\footnote{This implies that the heterotic construction based
on $\Phi$, in which such cubic terms are present (see Eq. (\ref{h})),
cannot be dual to a type II,
$Z_2 \times Z_2$ symmetric orbifold, for which the intersection matrix
is trivial.},
as in our case. On the other hand, the absence of
constant term both in $h^{(1)}$ and $\tilde{h}^{(1)}$ reflects the
vanishing of the Euler characteristic $\chi=2\left(
h^{1,1}-h^{2,1}\right)$.

\subsection{\sl Non-perturbative contributions}

Let us now try to go beyond the above perturbative result. We will
concentrate on the construction based on $\tilde{\Phi}$, for which we have
been able to determine the precise duality map. For the prepotential,
however, we have no exact type II result that could be used to
obtain the heterotic non-perturbative contributions directly.
Nevertheless, the structure of the type II model is useful to infer at
least part of these contributions. 

The type II $(1,1)$ ground state of
Section 2 is symmetric under permutations of the three moduli $\{T^1,
-1/T^2, T^3\}$. 
The heterotic dual should therefore possess the same
property with respect to 
$\{\tau_{S_{\rm Het}}, -1/T,U\}$. In fact,  invariance under
$-1/T \leftrightarrow U$ is a residual target-space duality symmetry
of the shifted lattices we are
considering, and Eq. (\ref{B}) is indeed invariant owing to the
covariance property\footnote{The absence of pure power-like terms in
(\ref{cov}) reflects again the vanishing of the Euler characteristic and
intersection form of the dual symmetric type II construction.}
of the perturbative result 
$\tilde{h}^{(1)}(T,U)$ given in (\ref{htilde}): 
\be
\tilde{h}^{(1)}\left(-{1\over U},-{1\over T}\right)={1 \over T^2\, U^2}
\tilde{h}^{(1)}(T,U)
\label{cov}
\ee
(note, however, that $\tilde{h}^{(1)}(U,T)\neq\tilde{h}^{(1)}(T,U)$,
owing to the breakdown of the $T \leftrightarrow U$ symmetry).

In order to promote the above $-1/T \leftrightarrow U$
permutation symmetry to the level of the three
moduli, we must demand the following covariance properties for the full
prepotential $\tilde{h}\left(\tau_{S_{\rm Het}},T,U\right)$:
\ba
\tilde{h}\left(\thet,-{1\over U},-{1\over T}\right)&=&
{1 \over T^2\, U^2}
\tilde{h}\left(\thet,T,U\right)\nn \\
\tilde{h}\left(-{1\over T},-{1\over \thet},U\right)&=&
{1 \over \thet^2\, T^2}
\tilde{h}\left(\thet,T,U\right)\label{npcov} \\
\tilde{h}\left(U,T,\thet\right)&=&
\tilde{h}\left(\thet,T,U\right)\, ,\nn
\ea
which are fulfilled by the tree-level contribution (\ref{h0}). We must
therefore add two more terms to 
$\tilde{h}^{(1)}(T,U)$:
\be
\tilde{h}\left(\thet,T,U\right)= h^{(0)}+ 
\tilde{h}^{(1)}(T,U)+
\tilde{h}^{(1)}\left(T,\tau_{S_{\rm Het}}\right)+
 T^2\, U^2 \, \tilde{h}^{(1)}\left(-{1\over U},\tau_{S_{\rm Het}}\right)\,
.
\label{htsym}
\ee
These extra terms account for non-perturbative corrections and are
exponentially suppressed at large $S_{\rm Het}$.

The above covariant symmetrization, which we have been advocating in
order to determine non-perturbative corrections to the prepotential, 
does not exclude the possibility of having also a
series of exponentially suppressed terms with, in the arguments,
covariant-symmetric functions of  $\tau_{S_{\rm Het}},T$ and $U$,
in the sense of (\ref{npcov}).
Unfortunately, we
have no reason to rule out such non-perturbative contributions, nor
a method for computing them from the type II symmetric or asymmetric 
ground states. 

\section{Conclusions}

In this work we studied $N=2$ superstring ground
states obtained from the heterotic and type II ten-dimensional
superstrings through freely acting (asymmetric) orbifold
compactification.
The massless spectrum of all these models is the same. Besides
the $N=2$ supergravity multiplet, there are three vector multiplets and
four hypermultiplets. The construction presented here extends the work
of Ref. \cite{gkp},
where we analysed heterotic/type II duals with $N=2$ supersymmetries that have
$3+N_V$ vector multiplets and $4+N_H$ hypermultiplets with $N_V=N_H
\ne 0$.
Here, we presented three different $N=2$ models with $N_V=N_H = 0$ 
and  we verified the non-perturbative duality conjecture between them.
The models we have considered are the following:
(\romannumeral1) the heterotic construction with $N=(2,0)$
supersymmetry, based on
characters $\tilde \Phi$ of the $c=(0,16)$ conformal block, Eq.
(\ref{88fivt})  (this choice is equivalent to
a particular choice of discrete Wilson lines, reducing the number of
the vectors to three and that of the  hypers to four);
(\romannumeral2) the type IIA symmetric construction with $N=(1,1)$
supersymmetry, which corresponds to a self-mirror Calabi--Yau
compactification with Hodge numbers $h^{1,1}=h^{2,1}=3$;
(\romannumeral3) the type II asymmetric $N=(2,0)$ 
freely acting orbifold
compactification,
where the initial $N=(4,4)$ supersymmetry is spontaneously broken to
$N=(2,0)$.

The equivalence of the heterotic $N=(2,0)$,  type IIA $N=(1,1)$ and
$N=(2,0)$ was  verified for the 
corrections of a modified gravitational and gauge combination
associated with the operator $\bP_{\rm grav}^{\prime 2}$
introduced in
Section 4. This operator has the property of being regular in the
entire $(T,U)$ moduli space. 

In the duality relations between the constructions described above, the
heterotic vector moduli
$(\tau_{S_{\rm Het}},T,U)$ are mapped to the three
$h^{1,1}$ moduli $(T^1,T^2,T^3)$ of the symmetric type II, as well as
to the moduli of the asymmetric type II $(\tilde{\tau}_{S_{\rm As}}, 
T_{\rm As}, U_{\rm As})$,
where $\tilde{\tau}_{S_{\rm As}}=-1/\tau_{S_{\rm As}}$ is the inverse of
the asymmetric type II dilaton.  Thus, there is a weak--strong coupling
$S$-duality relation between the heterotic and
the asymmetric type II ground state, $\tau_{S_{\rm Het}}
=-1/\tau_{S_{\rm As}}$. In all above duals there is a
(non-)perturbative restoration of $N=8$ and $N=4$ supersymmetry in
some specific limits of the three moduli, which is in agreement
with the duality maps.

By using these duality maps,
we found that {\it the type II
corrections provide the complete, perturbative and non-perturbative,
heterotic corrections}, as was
also the case for all $N=2$ models  with $N_V=N_H$ constructed in
Ref. \cite{gkp}. This remarkable  property is due to the universality
of the $N=2$ sector in the heterotic orbifold and of the
corresponding $N=(2,2)$ sectors in the symmetric and asymmetric type
II orbifolds. We obtained in this way the full gravitational heterotic
corrections. These contain instanton corrections,
$n_k\,\exp {2k\pi i \tau_{S_{\rm Het}}}$, which are due
to the Euclidean five-brane wrapped around the six-dimensional internal
space; they depend only on $\tau_{S_{\rm Het}}$ and not on the other
moduli. The explicit expressions for these corrections are given in
Eq. (\ref{np}). The Olive--Montonen duality group is a $\Gamma(2)$
subgroup of $SL(2,Z)_{\tau_{S_{\rm Het}}}$.

Finally, we computed the perturbative 
and part of the non-perturbative corrections to the prepotential for the
heterotic ground state. 
The perturbative result  is actually valid beyond
the models presented in this paper, and covers more general situations,
with $N_V=N_H\neq 0$ such as those of Ref. \cite{gkp}. The
non-perturbative  piece
was reached by using the triality symmetry between the three moduli
of the type II symmetric model
$T^1\leftrightarrow -1/T^2\leftrightarrow T^3$. The requirement
of (partial) restoration
of the ($N=4$) $N=8$ supersymmetry in special limits of the 
vector moduli turned to be too weak to rule out or determine
extra potential non-perturbative terms in the prepotential.

\vskip 1.cm
\centerline{\bf Acknowledgements}
\noindent
The authors thank E. Kiritsis for valuable discussions.
P.M. Petropoulos acknowledges the hospitality of the CERN Theory Division. 
A. Gregori thanks the Swiss National Science Foundation and 
the Swiss Office for Education and Science (ofes 95.0856).
This work was partially supported by the EEC under the contracts
TMR-ERBFMRX-CT96-0045 and TMR-ERBFMRX-CT96-0090.

\vskip 0.3cm
\setcounter{section}{0}
\setcounter{equation}{0}
\renewcommand{\theequation}{A. \arabic{equation}}
\section*{\normalsize{\centerline{\bf Appendix:
The trilogarithm series}}}
\noindent
We quote here the explicit solution for the series ${\cal L}_{c,a,b}$
and ${\cal L}^{(\l)}_{c,a,b}(T,U)$
appearing in the expressions of the
one-loop prepotential (see 
Section 6).
We have:   
\ba
{\cal L}_{c}(T,U) &=&
\Li_3\left(
e^{2\pi i\left(T-U\right)}
\right)
\nonumber\\
&&+c_0 \sum_{k>0}\left(2\Li_3\left(e^{4\pi i
T k }\right)-\Li_3\left(e^{2\pi i
T k}\right)\right)
\nonumber\\
&&+c_0 \sum_{\ell>0}\left(2\Li_3\left(e^{4\pi i
U \ell }\right)-\Li_3\left(e^{2\pi i
U \ell}\right)\right)
\nonumber\\
&& 
+\sum_{k,\ell>0}\Big(
-c_{k\ell}\,\Lii{T k+U \ell}\nonumber\\
&&\ \ \ \ \ \ \ \ \
+2\,c_{4k\ell}\,\Lii{2 T k +2 U \ell}
\nonumber\\
&&\ \ \ \ \ \ \ \ \
+2\,c_{4k\ell-2k-2\ell+1}\Lii{T(2k-1)+U(2\ell-1)}\Big)
\label{Lc}
\ea
\ba
{\cal L}_{a}(T,U) &=&
\sum_{k,\ell>0}\bigg(
a_{4k\ell-3k-3\ell+2}\,\Lii{\frac{T}{2}(4k-3)+
\frac{U}{2}(4\ell-3)}\nonumber\\
&&\ \ \ \ \ \
+a_{4k\ell-k-\ell}\,\Lii{\frac{T}{2}(4k-1)+
\frac{U}{2}(4\ell-1)}
\bigg)
\label{La}
\ea
\ba
{\cal L}_{b}(T,U) &=& b_{-1}
\Li_3\left(
e^{2\pi i\left(\frac{T}{2}-\frac{U}{2}\right)}
\right)
\nonumber\\
&&+\sum_{k,\ell>0}\bigg(
b_{4k\ell-k-3\ell}\,\Lii{\frac{T}{2}(4k-3)+
\frac{U}{2}(4\ell-1)}\nonumber\\
&&\ \ \ \ \ \ \ \ \
+b_{4k\ell-3k-\ell}\,\Lii{\frac{T}{2}(4k-1)+
\frac{U}{2}(4\ell-3)}
\bigg)\, , 
\label{Lb}
\ea
where the coefficients in the above expansions are defined through
\ba
\frac{\Omega\oao{0}{1}}{\eta^{24}}&=&
\frac{1}{q} +
\sum_{n\ge 0}c_n\,q^n\nonumber\\
\frac{\Omega\oao{1}{0}+\Omega\oao{1}{1}}{\eta^{24}}&=&
\sum_{n\ge 0}a_n\,q^{n+\frac{1}{4}}\label{Oex}\\
\frac{\Omega\oao{1}{0}-\Omega\oao{1}{1}}{\eta^{24}}&=&
\sum_{n\ge -1}b_n\,q^{n+\frac{3}{4}}\, .\nonumber
\ea

We also have
\ba
{\cal L}^{(0)}_{c}(T,U)&=&
\Li_3
\left(e^{2\pi i
\left(T-U \right)}\right)
\nonumber \\
&&+c^{(0)}_0 \sum_{k>0}
\left(\Li_3
\left(e^{2\pi i
T k}
\right)
+ \Li_3
\left(e^{2\pi i
U k }
\right)
\right)
\nonumber \\
&&+\sum_{k,\ell>0}
c_{k\ell}^{(0)}\,\Li_3
\left(e^{2\pi i
(T k +U \ell)}\right)
\label{Lc0}
\ea
\ba  
{\cal L}_{a}^{(0)}\left(\frac{T}{2},2U\right) &=&
a^{(0)}_0 \sum_{k>0}\left(\Li_3\left(e^{\pi i
T k }\right)+\Li_3\left(e^{4\pi i
U k}\right)\right)
\nonumber\\
&&
+\sum_{k,\ell>0}
a_{k\ell}^{(0)}\,\Li_3
\left(e^{2\pi i
\left(\frac{T}{2} k +2 U \ell\right)}\right)
\label{La0}
\ea
\ba
{\cal L}_{b}^{(0)}\left(\frac{T}{2},U\right) &=&
288 \sum_{k>0}\left(\Li_3\left(e^{2\pi i
T k }\right)+\Li_3\left(e^{4\pi i
U k}\right)\right)
\nonumber\\
&&
+\sum_{k,\ell>0}\bigg(
\left(2\, c_{2k\ell}^{(0)}-a_{2k\ell}^{(0)}\right)
\Li_3\left(e^{2\pi i(T k +2 U \ell)}\right)\nonumber\\
&&\ \ \ \ \ \ \ \ \
+b_{2k\ell-k-\ell}^{(0)}\,\Li_3
\left(e^{2\pi i
\left(\frac{T}{2} (2k-1) + U (2 \ell-1)\right)}\right)
\bigg)\, ,
\label{Lb0}
\ea
and
\ba
{\cal L}_{c}^{(1)}(T,U) &=&
{\cal L}_{c^{(1)}}(T,U)
\label{L1c}\\
{\cal L}_{a}^{(1)}(T,U) &=&
{\cal L}_{a^{(1)}}(T,U)
\label{L1a}\\
{\cal L}_{b}^{(1)}(T,U) &=&
{\cal L}_{b^{(1)}}(T,U)\, ,
\label{L1b}\\
\ea
where ${\cal L}_{c^{(1)},a^{(1)},b^{(1)}}(T,U)$ are displayed in
(\ref{Lc})--(\ref{Lb}), and
 $c_n^{(\l)}$, $a_n^{(\l)}$ and $b_n^{(\l)}$
are given by
\ba
\frac{\Omega^{(\l)}\oao{0}{1}}{\eta^{24}}&=&
\frac{1}{q} +
\sum_{n\ge 0}c_n^{(\l)}\,q^n\nonumber\\
\frac{\Omega^{(\l)}\oao{1}{0}+\Omega^{(\l)}\oao{1}{1}}{\eta^{24}}&=&
\sum_{n\ge 0}a_n^{(\l)}\,q^{n+\frac{\l}{4}}\label{Oex1}\\
\frac{\Omega^{(\l)}\oao{1}{0}-\Omega^{(\l)}\oao{1}{1}}{\eta^{24}}&=&
\sum_{n\ge 0}b_n^{(\l)}\,q^{n+\frac{\l+2}{4}}\, .\nonumber
\ea

We finally recall that the non-trivial monodromy properties of the above
functions are due to the following connexion formula for the trilogarithm:
$$
\Li_3\left(e^{x}\right)=
\Li_3\left(e^{-x}\right)+
{\pi^2 \over 3}x-
{i\pi \over 2}x^2-
{1 \over 6}x^3 \sp {\rm for \ \ } \Re x \geq 0 
\, .
$$

\end{document}